\documentclass{PoS}
\pdfoutput=1

\usepackage{setspace}

\usepackage{amsmath}
\usepackage{mathtools}
\usepackage{epstopdf}

\usepackage{verbatim}
\usepackage{booktabs}
\usepackage{arydshln}
\usepackage{array}

\usepackage{color}
\usepackage{pstricks,framed}

\usepackage{tensor}
\usepackage{amssymb}
\usepackage{amsthm}
\usepackage{amsfonts}
\usepackage{simplewick}

\usepackage{chngcntr}
\usepackage[retainorgcmds]{IEEEtrantools}

\usepackage{graphicx}

\counterwithin{equation}{section}
\title{Large-Spin Expansions of Giant Magnons}
\author{\speaker{Georgios Linardopoulos}\\
Department of Physics, National and Kapodistrian University of Athens,\\
Zografou Campus, 157 84, Athens, Greece\\
Institute of Nuclear and Particle Physics, N.C.S.R., "Demokritos",\\
153 10, Agia Paraskevi, Greece \\
E-mail: \email{glinard@inp.demokritos.gr}}
\abstract{This is a talk delivered at the Workshop on Quantum Fields and Strings of the 2014 Corfu Summer Institute. We discuss how giant magnons emerge in the context of the AdS$_5$/CFT$_4$ correspondence as the gravity duals of $\mathcal{N} = 4$ super Yang-Mills magnon excitations. Then we present a new analytic expression for the dispersion relation of classical finite-size giant magnons with Lambert's W-function.}
\FullConference{Proceedings of the Corfu Summer Institute 2014 \\
3-21 September 2014\\
Corfu, Greece}
\ShortTitle{Large-Spin Expansions of Giant Magnons}
\begin{document}
\maketitle\flushbottom\normalsize
\newpage
\tableofcontents
\section[Introduction and Motivation]{Introduction and Motivation \label{Section:Introduction}}
\noindent Perhaps the most astounding prediction of the AdS/CFT correspondence \cite{Maldacena97, GubserKlebanovPolyakov98, Witten98a},
\begin{equation}
\mathcal{N} = 4,\ \mathfrak{su}\left(N_c\right)\ \text{super Yang-Mills theory} = \text{IIB superstring theory on AdS}_5\times\text{S}^5, \label{AdS-CFT_Correspondence}
\end{equation}
is that the observables of $\mathcal{N} = 4$ super Yang-Mills (SYM) theory and IIB string theory on AdS$_5\times\text{S}^5$ (spectra, correlation functions, scattering amplitudes, Wilson loops, etc.) can be put in one-to-one correspondence. That is, for each and every observable of $\mathcal{N} = 4$ SYM there exists a dual and equal observable of IIB string theory on AdS$_5\times\text{S}^5$. In the years that have passed since the formulation of AdS/CFT, a multitude of non-trivial checks (e.g.\ between symmetries, spectra, correlation functions, anomalies, etc.\footnote{The interested reader is referred to \cite{MAGOO99} for an early but complete discussion of AdS/CFT tests.}) has been performed, allowing to verify the validity of the duality and elucidate the way that the mapping between the properties of the two implicated theories works. \\[6pt]
\indent The need for a "dictionary" of AdS/CFT is most pronounced when we try to match the spectra of the two theories, which is their most important observable. Since conformal field theories (CFTs) do not have asymptotic states/particles, it is not exactly clear what the spectrum of $\mathcal{N} = 4$ SYM (which is a CFT) must be. According to the field/operator correspondence of AdS/CFT however, it is local gauge-invariant operators that play the role of particles in $\mathcal{N} = 4$ SYM and their scaling dimensions compose the gauge theory's spectrum. This spectrum must then match the one of IIB string theory on AdS$_5\times\text{S}^5$ that is made up from the string state energies. The matching of spectra in AdS/CFT correspondence generally proceeds according to the following plan: \\[12pt]
\indent 1. Compute the scaling dimensions $\Delta$ of all gauge-invariant operators of $\mathcal{N} = 4$ SYM. \\[6pt]
\indent 2. Compute the energies $E$ of IIB superstring states in AdS$_5\times\text{S}^5$. \\[6pt]
\indent 3. Map the operators of $\mathcal{N} = 4$ SYM to IIB string states in AdS$_5\times\text{S}^5$. \\[6pt]
\indent 4. Compare the operator dimensions $\Delta$ with the dual string energies $E$ and find agreement. \\[12pt]
\indent This looks like a gargantuan program however and it must be broken down into many smaller and doable parts. One obvious simplification restricts our attention to (local) single-trace operators that are dual to single-particle states. Secondly, we usually consider the planar limit in which the number of colors on the gauge theory side becomes infinite ($N_c \rightarrow \infty$) and string theory becomes free ($g_s \rightarrow 0$). Thirdly, we often focus on the various closed sectors of AdS/CFT and examine certain classes of its states and operators. One such class is formed by BPS or chiral primary operators that are annihilated by one or more of the Poincar\'{e} supercharges and are protected from receiving quantum corrections. They are dual to free point-like strings of IIB string theory. Another sector where the spectra of AdS/CFT have been found to agree is the Berenstein-Maldacena-Nastase (BMN) sector \cite{BMN02} consisting of 'almost' BPS operators, dual to 'nearly' point-like free string states. \\[6pt]
\indent Beyond the BPS and BMN limits, all tests point out that there's a perfect match between the spectra of AdS/CFT, at least as far as the planar/free string limit is concerned. Actually the planar limit of AdS/CFT is interesting for one more reason: both theories are thought to be quantum integrable in this limit. \\[6pt]
\indent Planar integrability has very profound consequences for AdS/CFT. In classical terms, a theory is integrable when it possesses the maximum allowed number of conservation laws that may in turn be integrated and the theory be solved. Indeed, it is claimed \cite{Beisertetal12} that integrability completely solves the spectral problem of AdS/CFT in the planar limit, in the sense that it provides the full set of algebraic equations that determine it. The fact that the planar spectra of both theories are determined by a common set of equations implies that they must match. Integrability also provides the computational toolkit for solving the planar AdS/CFT, i.e.\ for computing all of its observables. \\[6pt]
\indent Integrability-based methods (e.g.\ TBA/Y-system/QSC\footnote{The acronyms TBA and QSC stand for thermodynamic Bethe ansatz and quantum spectral curve respectively.}) do have their limitations. There exist regimes of AdS/CFT where solving the system of algebraic equations that determines the spectrum becomes so cumbersome that it is almost impossible to tackle it either computationally or analytically. For the spectra of long $\mathcal{N} = 4$ SYM operators at strong coupling that are dual to long semiclassical strings, e.g.\ GKP strings \cite{GubserKlebanovPolyakov02} and giant magnons \cite{HofmanMaldacena06}, the input from integrability is still rather poor. There are various reasons why we want to actually be able to compute the planar AdS/CFT spectrum. First and foremost, the scope of AdS/CFT becomes somewhat limited if we do not know how to compute its full spectrum. Secondly, besides just wanting to check the matching of the spectra explicitly, we also want to complete the AdS/CFT dictionary via the state/operator correspondence that we saw above. Thirdly, we would like to explore the possibility of finding closed-form expressions in the AdS/CFT spectrum. \\[6pt]
\indent In a recent paper \cite{FloratosLinardopoulos14}, we introduced a method to calculate the classical spectrum of giant magnons without using integrability. The results of this paper have not been obtained previously with any other method. Being semianalytical, it is also impossible to obtain them by means of a computer. Developing a spectral method that is not based on integrability has the advantage of being applicable even in those cases where integrability becomes too difficult to handle. Furthermore, such methods take us beyond the idealized integrable paradigms (e.g.\ non-planar AdS/CFT, QCD, p-branes) and may allow us to compute the spectra in more generic frameworks. We shall also see that we can go a long way towards finding closed formulas in the AdS/CFT spectrum. \\[6pt]
\indent This paper is the written version of a homonymous talk\footnote{The slides of the talk can be found in the address \href{http://www.physics.ntua.gr/corfu2014/lectures.html}{http://www.physics.ntua.gr/corfu2014/lectures.html}.} delivered at the 2014 Corfu Summer Institute. It is based on the paper \cite{FloratosLinardopoulos14} and it is organized in two main parts. In \S\ref{Section:InfiniteSizeGMs} we introduce infinite-size, aka Hofman-Maldacena (HM) giant magnons and discuss their emergence in the context of the AdS/CFT correspondence as the string theory duals of $\mathcal{N} = 4$ SYM magnon excitations. In \S\ref{Section:FiniteSizeGMs} we discuss the finite-size generalization of giant magnons and present a new analytic expression for their classical dispersion relation with Lambert's W-function. In the discussion section \S\ref{Section:Discussion}, we summarize our work and give a list of some interesting future projects.
\section[Infinite-Size Magnons]{Infinite-Size Magnons \label{Section:InfiniteSizeGMs}}
\subsection[$\mathcal{N} = 4$ SYM Magnons]{$\mathcal{N} = 4$ SYM Magnons \label{Subsection:Magnons}}
\noindent Let us now briefly see how the concept of the magnon emerges in $\mathcal{N} = 4$ SYM theory. As we have explained above, in a conformal field theory like $\mathcal{N} = 4$ SYM, it is operators that take the role of particles and the spectrum is formed by the operator scaling dimensions. We will focus on the $\mathfrak{su}\left(2\right)$ sector of $\mathcal{N} = 4$ SYM which consists of the following single-trace operators:
\begin{equation}
\mathcal{O}^{\left(J,M\right)} = \text{Tr}\left[\mathcal{Z}^{J}\mathcal{X}^{M}\right] + \ldots, \quad L \equiv J + M, \label{Operators_su(2)}
\end{equation}
where $\mathcal{X}, \ \mathcal{Y}, \ \mathcal{Z}$ are the three complex scalar fields of $\mathcal{N} = 4$ SYM, composed out of the six real scalars $\phi_i$ of the theory ($i = 1,2,\ldots,6$). The dots in \eqref{Operators_su(2)} stand for all possible permutations of the fields inside the trace, while each term in the sum \eqref{Operators_su(2)} must be multiplied by a suitable coefficient (which we omit for simplicity). \\[6pt]
\indent Due to the cyclic property of the trace in \eqref{Operators_su(2)}, we may regard the complex fields $\mathcal{Z}$ as the ground state fields (spin up) and $\mathcal{X}$ as some sort of impurities (spin down) in a closed spin chain. The length of the spin chain is $L$, while $J$ is its spin and $M$ is the number of magnons. E.g.\ one permutation of a spin chain with $\left(L,J,M\right) = \left(13,8,5\right)$ is
\begin{center}\begin{tabular}{m{3cm}m{1cm}m{4.5cm}m{.5cm}m{3cm}}
\centering{$\text{Tr}\left[\mathcal{Z}^5\mathcal{X}^2\mathcal{Z}^3\mathcal{X}^3\right]$} & \centering{$\longleftrightarrow$} & \centering{\includegraphics[scale=.5]{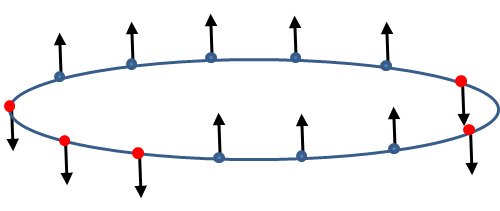}} & \centering{=} & \centering{$|\uparrow\uparrow\uparrow\uparrow\uparrow\downarrow\downarrow\uparrow\uparrow\uparrow\downarrow\downarrow\downarrow\rangle$.}
\end{tabular}\end{center}

\indent It was proven in 2002 by Minahan and Zarembo \cite{MinahanZarembo03} that at one loop, the dilatation operator $\mathbb{D}$ of the $\mathfrak{su}\left(2\right)$ sector \eqref{Operators_su(2)} of $\mathcal{N} = 4$ SYM is given by the Hamiltonian of the Heisenberg XXX$_{1/2}$ quantum spin chain:
\begin{IEEEeqnarray}{c}
\mathbb{D} = L \cdot \mathbb{I} + \frac{\lambda}{8\pi^2}\,\mathbb{H} + \sum_{n = 2}^\infty \lambda^n \mathbb{D}_n, \quad \mathbb{H} = \sum_{j = 1}^L \left(\mathbb{I}_{j,j + 1} - \mathbb{P}_{j,j + 1}\right) = 2\sum_{j = 1}^{L}\left(\frac{1}{4} - \textbf{S}_{j}\cdot \textbf{S}_{j+1}\right), \ \textbf{S} \equiv\frac{\boldsymbol\sigma}{2}, \qquad \label{SpinChainHamiltonian}
\end{IEEEeqnarray}
where $\lambda \equiv g^2_{YM}N_c$ is the 't Hooft coupling, $\boldsymbol\sigma$ are the Pauli matrices and the indices $j$, $j + 1$ in \eqref{SpinChainHamiltonian} indicate that the corresponding matrix acts only on the positions $j$ and $j + 1$. $\mathbb{I}_{i,j}$ and $\mathbb{P}_{i,j}$ are the spin-identity and spin-exchange operators:
\begin{IEEEeqnarray}{c}
\left(\mathbb{I}_{i,j}\right)_{abcd} \equiv \left(\delta_{ab}\right)_i\left(\delta_{cd}\right)_j, \qquad \mathbb{P}_{i,j} \equiv \frac{1}{2}\left(\mathbb{I}_{i,j} + \boldsymbol\sigma_i \cdot \boldsymbol\sigma_j\right).
\end{IEEEeqnarray}
\indent The Heisenberg XXX$_{1/2}$ spin chain can be diagonalized by the (coordinate) Bethe ansatz (BA). Without going into too many details (the reader is referred to the review \cite{Plefka05} for a complete discussion), the eigenvalues of the dilatation operator \eqref{SpinChainHamiltonian} that correspond to M-magnon operators
\begin{IEEEeqnarray}{l}
\text{Tr}\left[\mathcal{Z}^{J}\mathcal{X}^M\right] \sim \left|x_1,x_2,\ldots,x_M\right\rangle = |\uparrow \ldots \uparrow{\color{red}\underset{x_1}{\downarrow}}\uparrow \ldots \uparrow{\color{red}\underset{x_2}{\downarrow}}\uparrow{\color{red} \ldots} \uparrow{\color{red}\underset{x_M}{\downarrow}}\uparrow \ldots \uparrow\rangle, \label{BetheEigenvectors}
\end{IEEEeqnarray}
are given by:
\begin{IEEEeqnarray}{l}
\Delta = J + M + \frac{\lambda}{2\pi^2}\sum_{j = 1}^M \sin^2\frac{p_j}{2} + O\left(\lambda^2\right), \qquad \sum_{j = 1}^M p_j = 0, \label{BetheEigenvalues}
\end{IEEEeqnarray}
where the vanishing of the total momentum follows from the cyclicity of the trace in \eqref{Operators_su(2)}. \\[6pt]
\indent To account for higher-loop contributions $\mathbb{D}_n$ to the dilatation operator \eqref{SpinChainHamiltonian}, Beisert, Dippel and Staudacher (BDS) proposed an all-loop, asymptotic Bethe ansatz (ABA) \cite{BeisertDippelStaudacher04}:
\begin{IEEEeqnarray}{c}
\Delta = J + M + \frac{\lambda}{8\pi^2} \sum_{j = 1}^{M} E\left(p_j\right)\;, \quad E\left(p_j\right) = \frac{8\pi^2}{\lambda}\left[\sqrt{1 + \frac{\lambda}{\pi^2}\sin^2\frac{p_j}{2}} - 1\right], \quad j = 1,2,\ldots,M. \qquad \label{AsymptoticBetheAnsatz}
\end{IEEEeqnarray}
The ansatz \eqref{AsymptoticBetheAnsatz} is asymptotic in the sense that there's a \textit{critical} loop order equal to the length of the spin-chain $L$ at which it stops being valid. At the critical loop order $L$, the range of the spin chain interactions becomes greater than the length of the chain and the so-called wrapping corrections have to be added to the dispersion relation \eqref{AsymptoticBetheAnsatz}. The wrapping corrections actually originate from higher genus corrections to the dilatation operator that we have neglected in the planar limit. From the string theory point of view, wrapping effects arise because of the finite circumference of the cylindrical worldsheet. \\[6pt]
\indent Let us consider $M = 1$ magnon states:\footnote{We note here that one-magnon operators with non-vanishing momentum $p$ do not correspond to physical states of the theory since as we saw, the trace condition \eqref{BetheEigenvalues} implies that their momentum must identically vanish. To accommodate single-magnon states, the corresponding symmetry algebra $\mathfrak{su}\left(2|2\right) \oplus \mathfrak{su}\left(2|2\right) \subset \mathfrak{psu}\left(2,2|4\right)$ must be extended with two central charges.\label{Footnote:SingleMagnonStates}}
\begin{IEEEeqnarray}{c}
\mathcal{O}_M = \sum_{m = 1}^{J+1} e^{imp} \left|\mathcal{Z}^{m-1}\mathcal{X}\mathcal{Z}^{J-m+1}\right\rangle, \quad p \in \mathbb{R}. \label{One-MagnonOperators}
\end{IEEEeqnarray}
At infinite size\footnote{In this paper $E,J = \infty, \ \omega = 1$ denotes infinite size (obtained by computing the limits $\lim_{J \rightarrow \infty,\omega \rightarrow 1}$), while $E,J \rightarrow \infty, \ \omega \rightarrow 1$ denotes large but still finite size.} $J = \infty$ there are no wrapping corrections and the corresponding BDS dispersion relation \eqref{AsymptoticBetheAnsatz} becomes exact to all-loops:
\begin{IEEEeqnarray}{l}
\Delta - J = \sqrt{1 + \frac{\lambda}{\pi^2}\sin^2\frac{p}{2}}\,, \quad J = \infty, \ \text{all } \lambda. \label{Magnon1}
\end{IEEEeqnarray}
It has been proven by Beisert in \cite{Beisert05b} that this relation follows by extending the corresponding symmetry algebra $\mathfrak{su}\left(2|2\right) \oplus \mathfrak{su}\left(2|2\right) \subset \mathfrak{psu}\left(2,2|4\right)$. We may obtain its weak and strong coupling limits as follows:
\begin{IEEEeqnarray}{ll}
\Delta - J = 1 + \frac{\lambda}{2\pi^2}\sin^2\frac{p}{2} - \frac{\lambda^2}{8\pi^4}\sin^4\frac{p}{2} + \frac{\lambda^3}{16\pi^6}\sin^6\frac{p}{2} - \ldots\,, \qquad & \lambda \rightarrow 0 \quad \text{(weak coupling)} \qquad \label{Magnon2} \\[12pt]
\Delta - J = \frac{\sqrt{\lambda}}{\pi}\sin\frac{p}{2} + 0 + \frac{\pi}{2\sqrt{\lambda}}\csc\frac{p}{2} - \frac{\pi^3}{8\lambda^{3/2}}\csc^3\frac{p}{2} + \ldots\,, \qquad &  \lambda \rightarrow \infty \quad \text{(strong coupling)}. \qquad \label{Magnon3}
\end{IEEEeqnarray}

\subsection[Hofman-Maldacena Giant Magnons]{Hofman-Maldacena Giant Magnons \label{Subsection:HM-GiantMagnons}}
\begin{figure}
\begin{center}
\includegraphics[scale=0.25]{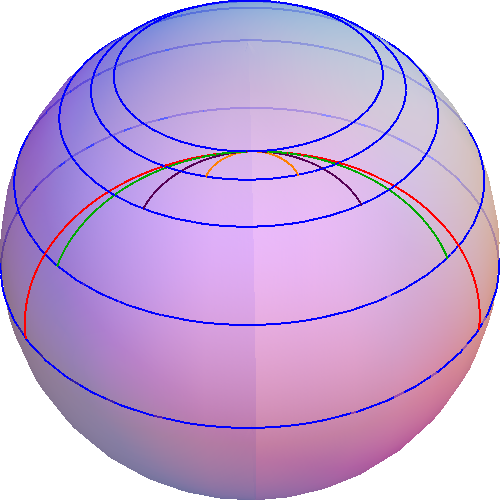} \hspace{2cm} \includegraphics[scale=0.25]{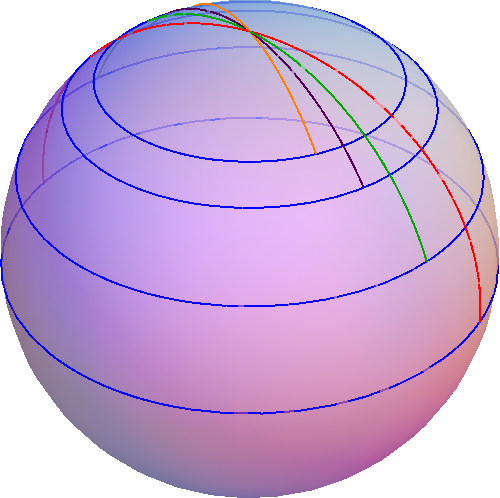}
\caption{Plots of giant magnons (left) and $\mathbb{R}\times\text{S}^2$ GKP strings (right) for various values of their angular velocity $\omega \geq 1$. Finite-size GMs ($v \neq 0$, $\omega \neq 1$) perform a wave-like motion around the 2-sphere. The GKP strings ($v = 0$) rotate rigidly around their fixed polar point. The infinite-size limits ($\omega = 1$) have been drawn with red color in both cases.} \label{Graph:GMs_and_GKPStringsII}
\end{center}
\end{figure}

\noindent The string theory duals of magnon operators \eqref{One-MagnonOperators} are the giant magnons (GMs). Giant magnons were found in 2006 by Hofman and Maldacena \cite{HofmanMaldacena06} and are open, single-spin strings that rotate rigidly in $\mathbb{R}\times\text{S}^2 \subset \text{AdS}_5 \times \text{S}^5$. Let the line element of AdS$_5 \times \text{S}^5$ be
\begin{IEEEeqnarray}{ll}
ds^2 = R^2 \Big[-\cosh^2\rho \, dt^2 &+ d\rho^2 + \sinh^2\rho \, \Big(d\overline{\theta}^2 + \sin^2\overline{\theta} \, d\overline{\phi}_1^2 + \cos^2\overline{\theta} \, d\overline{\phi}_2^2\Big) + \nonumber \\[6pt]
& + d\theta^2 + \sin^2\theta \, d\phi^2 + \cos^2\theta \, \left(d\theta_1^2 + \sin^2\theta_1 \, d\phi_1^2 + \cos^2\theta_1 \, d\phi_2^2\right)\Big]. \qquad
\end{IEEEeqnarray}
Then, the HM giant magnon is described by the following ansatz:
\begin{IEEEeqnarray}{c}
\Big\{t = \tau, \rho = \overline{\theta} = \overline{\phi}_1 = \overline{\phi}_2 = 0\Big\} \times \Big\{\theta = \theta\left(\sigma - v\tau\right), \phi = \tau + \varphi\left(\sigma - v\tau\right), \theta_1 = \phi_1 = \phi_2 = 0\Big\}, \qquad \; \label{GiantMagnonAnsatz1}
\end{IEEEeqnarray}
where $0 \leq |v| \leq 1$ is the linear velocity of the GM. Physically, a HM giant magnon of (conserved) linear momentum $p$ corresponds to an arc of (constant) angular extent $\Delta\phi = p$ that extends between the equator and the parallel $\zeta_v$\,:
\begin{IEEEeqnarray}{c}
0 \leq z \leq \zeta_v \leq R, \quad \zeta_v \equiv R\sqrt{1 - v^2}, \quad z \equiv R\cos\theta.
\end{IEEEeqnarray}
\indent The HM giant magnon has been drawn with red color on the left sphere of figure \ref{Graph:GMs_and_GKPStringsII}. HM giant magnons have infinite size since their conserved charges (which measure their "size") both diverge, $E, J = \infty$.\footnote{The epithet "giant" derives from the fact that (both finite and infinite-size) giant magnons are "long" strings ($E, J \rightarrow \infty$) that "see" the curvature of the 2-sphere upon which they live. Conversely, the motion of "short" strings ($E, J \rightarrow 0$) takes place in an almost flat background. See figure \ref{Graph:GM_Momentum-Spin-Energy}: long strings correspond to $\omega \rightarrow 1$, while for short strings $\omega \rightarrow \infty$.} However their difference remains finite:
\begin{IEEEeqnarray}{c}
E - J = \frac{\sqrt{\lambda}}{\pi} \, \left|\sin\frac{\Delta\varphi}{2}\right|, \qquad J = \infty, \ \sqrt{\lambda} = \frac{R^2}{\alpha'} \rightarrow \infty, \label{GiantMagnon1}
\end{IEEEeqnarray}
which is nothing more than the classical part (tree level) of the strong coupling limit \eqref{Magnon3} of the 1-magnon BDS dispersion relation \eqref{Magnon1}. \\[6pt]
\indent By using a duality between classical strings in $\mathbb{R}\times\text{S}^2$ and classical sine-Gordon solitons that is known as Pohlmeyer reduction \cite{Pohlmeyer75}, Hofman and Maldacena also showed that the scattering matrix of GMs coincides with the strong-coupling limit of the gauge theory prediction \cite{ArutyunovFrolovStaudacher04}. The upshot is that infinite-size giant magnons are dual to the one-magnon states \eqref{One-MagnonOperators} of $\mathcal{N} = 4$ SYM, $\left|\ldots\mathcal{Z} \mathcal{Z}\mathcal{X}\mathcal{Z}\mathcal{Z}\ldots\right\rangle \sim |\ldots \uparrow\uparrow{\color{red}\downarrow}\uparrow\uparrow \ldots\rangle$. \\[6pt]
\indent As we have already noted in footnote \ref{Footnote:SingleMagnonStates}, single-magnon states with non-vanishing momentum $p$ are incompatible with the trace condition \eqref{BetheEigenvalues}, according to which the total magnon momentum should vanish. Likewise, open string states like giant magnons are incompatible with the spectrum of type IIB superstring theory which contains only closed strings. In order to obtain meaningful configurations on both sides of AdS/CFT, two or more (giant) magnons must be superposed so as to form closed string states with vanishing total momentum.  \\[6pt]
\indent Superimposing two giant magnons with velocity $v = 0$, maximum angular extent $\Delta\varphi = \pi$ and angular momenta $J/2$, gives rise to the Gubser-Klebanov-Polyakov (GKP) string in $\mathbb{R}\times\text{S}^2$ \cite{GubserKlebanovPolyakov02}:
\begin{IEEEeqnarray}{c}
\Big\{t = \tau, \rho = \overline{\theta} = \overline{\phi}_1 = \overline{\phi}_2 = 0\Big\} \times \Big\{\theta = \theta\left(\sigma\right), \phi = \tau, \theta_1 = \phi_1 = \phi_2 = 0\Big\}, \label{GKP_String_II1}
\end{IEEEeqnarray}
the dispersion relation of which at infinite size ($E, J = \infty$) is:
\begin{IEEEeqnarray}{c}
E - J = \frac{2\sqrt{\lambda}}{\pi}, \qquad J = \infty, \ \lambda \rightarrow \infty. \label{AnomalousDimensions12}
\end{IEEEeqnarray}
The $\mathbb{R}\times\text{S}^2$ GKP string \eqref{GKP_String_II1} is dual to the 2-magnon operator $\text{Tr}\left[\mathcal{Z}^J\mathcal{X}^2\right]$ of $\mathcal{N} = 4$ SYM. It is depicted with red color on the right sphere of figure \ref{Graph:GMs_and_GKPStringsII}. GKP strings are closed folded strings that rotate rigidly about their fixed polar points on the 2-sphere.
\section[Finite-Size Giant Magnons]{Finite-Size Giant Magnons \label{Section:FiniteSizeGMs}}
\subsection[Finite-Size Giant Magnons]{Finite-Size Giant Magnons \label{Subsection:GMs}}
\begin{figure}
\begin{center}
\includegraphics[scale=0.25]{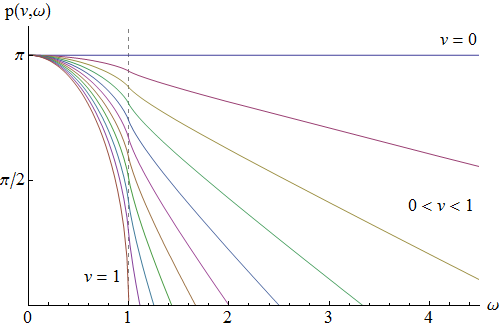}
\
\includegraphics[scale=0.25]{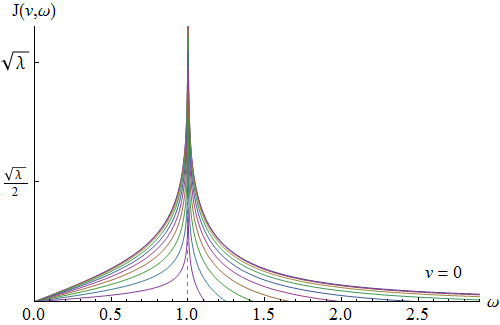}
\
\includegraphics[scale=0.25]{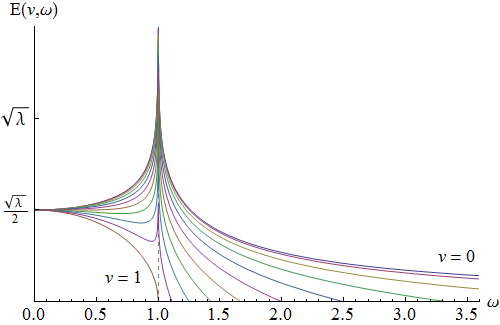}
\caption{Momentum, spin and energy of the giant magnon as functions of its angular velocity $\omega$.} \label{Graph:GM_Momentum-Spin-Energy}
\end{center}
\end{figure}

\noindent The finite-size generalization of the giant magnon can be obtained from the following ansatz:
\begin{IEEEeqnarray}{c}
\Big\{t = \tau, \rho = \overline{\theta} = \overline{\phi}_1 = \overline{\phi}_2 = 0\Big\} \times \Big\{\theta = \theta\left(\sigma - v\omega\tau\right), \phi = \omega\tau + \varphi\left(\tau,\sigma\right), \theta_1 = \phi_1 = \phi_2 = 0\Big\}, \qquad \label{GiantMagnonAnsatz2}
\end{IEEEeqnarray}
where $v$ is the string's linear velocity and $\omega$ is its angular velocity. \\[6pt]
\indent Depending on the relative values of the velocities $v$ and $\omega$, there exist two basic configurations of \eqref{GiantMagnonAnsatz2}, namely giant magnons (for which $v \cdot \omega \leq 1$) and single spikes ($v \cdot \omega \geq 1$) each of which contains two different sub-domains, the elementary and the doubled. Of these, only the elementary GMs are stable while the doubled GMs and the single spikes (elementary or doubled) are unstable. For a more thorough discussion the reader is referred to the paper \cite{FloratosLinardopoulos14}. In this talk, we will mainly focus on the stable elementary region of giant magnons (in which $0 \leq |v| \leq 1/\omega \leq 1$), although we will see that our results can be simply extended to the doubled region ($0 \leq |v| \leq 1 \leq 1/\omega$). \\[6pt]
\indent Contrary to the HM giant magnons, finite-size elementary GMs\footnote{From now on and unless otherwise noted, the term giant magnon will exclusively refer to the stable GMs of the elementary region, for which $0 \leq |v| \leq 1/\omega \leq 1$.} do not touch the equator of the 2-sphere, but extend between the parallels $\zeta_{\omega}$ and $\zeta_v$:
\begin{IEEEeqnarray}{c}
0 \leq \zeta_{\omega} \leq z \leq \zeta_v \leq R, \qquad \zeta_{\omega} \equiv R\sqrt{1 - \frac{1}{\omega^2}}, \quad \zeta_v \equiv R\sqrt{1 - v^2}, \quad z \equiv R\cos\theta.
\end{IEEEeqnarray}
\indent The finite-size GM has been plotted for various values of its angular velocity $\omega$ on the left sphere of figure \ref{Graph:GMs_and_GKPStringsII}. The red-colored giant magnon corresponds to the HM magnon of infinite size. Finite-size GMs perform a wave-like (or "worm-like") revolution around the 2-sphere. They still have three conserved charges, namely their energy $E$, spin $J$ and momentum/angular extent $p = \Delta\phi$. The GM charges have been plotted as functions of the angular velocity $\omega$ and various values of the linear velocity $0 \leq v \leq 1$ in both their elementary ($\omega \geq 1$) and doubled ($\omega \leq 1$) regions in figure \ref{Graph:GM_Momentum-Spin-Energy}. \\[6pt]
\indent Infinite-size giant magnons with $E,J = \infty$ are recovered in the limit $\omega = 1$ (cf.\ figure \ref{Graph:GM_Momentum-Spin-Energy}), in which the magnon's elementary and doubled regions merge into the HM region, $0 \leq |v| \leq 1$ and rigid body motion is restored. By superposing two finite-size GMs with velocity $v = 0$, maximum momentum/angular extent $p = \Delta\phi = \pi$ and angular momentum equal to $J/2$, we obtain the finite-size version of the GKP string \eqref{GKP_String_II1}:
\begin{IEEEeqnarray}{c}
\Big\{t = \tau, \rho = \overline{\theta} = \overline{\phi}_1 = \overline{\phi}_2 = 0\Big\} \times \Big\{\theta = \theta\left(\sigma\right), \phi = \omega\tau, \theta_1 = \phi_1 = \phi_2 = 0\Big\}. \label{GKP_String_II2}
\end{IEEEeqnarray}
The ansatz \eqref{GKP_String_II2} follows from \eqref{GiantMagnonAnsatz2} by using the formulas that can be found in appendix A1 of \cite{FloratosLinardopoulos14}. As we have already said, GKP strings in $\mathbb{R}\times\text{S}^2$ are dual to the 2-magnon operators $\text{Tr}\left[\mathcal{Z}^J\mathcal{X}^2\right]$ of $\mathcal{N} = 4$ SYM. At finite-size, the 2-magnon operators $\text{Tr}\left[\mathcal{Z}^J\mathcal{X}^2\right]$ have large yet finite length $L = J + 2$. The finite-size GKP string in $\mathbb{R}\times\text{S}^2$ has been drawn for various values of the angular velocity $\omega \geq 1$ on the right sphere of figure \ref{Graph:GMs_and_GKPStringsII}. Red color corresponds to the infinite-size case for which $\omega = 1$.
\subsection[Dispersion Relation]{Dispersion Relation \label{Subsection:GM-DispersionRelation}}
\noindent Ideally, we would be able to write down an exact all-loop dispersion relation for finite-size giant magnons just as we did in the case of the Hofman-Maldacena giant magnon with the all-loop formula \eqref{Magnon1}. Unfortunately this seems to be a very complicated problem. The general form of the dispersion relation of finite-size GMs and equivalently finite-size, one-magnon states \eqref{One-MagnonOperators} of $\mathcal{N} = 4$ SYM at strong coupling is the following:
\begin{IEEEeqnarray}{c}
E - J = \epsilon_{\infty} + \underset{\text{finite-size corrections}}{\underbrace{\sqrt{\lambda}\,\delta\epsilon_{\text{cl}} + \delta\epsilon_{1\text{-loop}} + \frac{1}{\sqrt{\lambda}}\delta\epsilon_{2\text{-loop}} + \ldots}}\,, \qquad J,\lambda \rightarrow \infty, \label{GiantMagnon2}
\end{IEEEeqnarray}
where $\epsilon_{\infty}$ is just the all-loop, 1-magnon formula \eqref{Magnon1},
\begin{IEEEeqnarray}{l}
\epsilon_{\infty} = \sqrt{1 + \frac{\lambda}{\pi^2}\sin^2\frac{p}{2}} = \frac{\sqrt{\lambda}}{\pi}\sin\frac{p}{2} + 0 + \frac{\pi}{2\sqrt{\lambda}}\csc\frac{p}{2} - \frac{\pi^3}{8\lambda^{3/2}}\csc^3\frac{p}{2} + \ldots, \label{GiantMagnon3}
\end{IEEEeqnarray}
to which \eqref{GiantMagnon2} reduces at infinite size $J = \infty$. At finite size, $\epsilon_{\infty}$ receives classical corrections $\delta\epsilon_{\text{cl}}$ and quantum (i.e.\ $\alpha'$ or $\lambda$) corrections $\delta\epsilon_{n\text{-loop}}$. By studying the motion of classical strings in $\mathbb{R}\times\text{S}^2$, Arutyunov, Frolov and Zamaklar \cite{ArutyunovFrolovZamaklar06} computed the first few terms of classical finite-size corrections $\delta\epsilon_{\text{cl}}$:\footnote{See also \cite{AstolfiForiniGrignaniSemenoff07}.}
\begin{IEEEeqnarray}{ll}
\delta\epsilon_{\text{cl}} = - \frac{4}{\pi} \,& \sin\frac{p}{2} \, \Bigg\{\sin^2\frac{p}{2}\,e^{- \mathcal{L}} + \bigg[8 \cos^2\frac{p}{2}\mathcal{J}^2 + 4\sin\frac{p}{2}\left(3\cos p + 2\right)\mathcal{J} + \nonumber \\[6pt]
& + \sin^2\frac{p}{2}\left(6\cos p + 7\right)\bigg]e^{- 2\mathcal{L}} + \ldots\Bigg\}, \quad \mathcal{J} \equiv \frac{\pi J}{\sqrt{\lambda}}, \quad \mathcal{L} \equiv 2\mathcal{J} \csc\frac{p}{2} + 2. \qquad \label{GiantMagnon4}
\end{IEEEeqnarray}
Many more terms of \eqref{GiantMagnon4} can be computed from classical strings with $\mathsf{Mathematica}$ (see e.g.\ the appendix B of \cite{FloratosLinardopoulos14}). Alternatively, the leading term of \eqref{GiantMagnon4} has been determined by finite-gap methods and the L\"{u}scher formulae \cite{JanikLukowski07, MinahanSax08, HellerJanikLukowski08}. Some terms of the leading quantum finite-size corrections $\delta\epsilon_{1\text{-loop}}$ have been computed by Gromov, Sch\"{a}fer-Nameki and Vieira in \cite{GromovSchafer-NamekiVieira08a, GromovSchafer-NamekiVieira08b}. \\[6pt]
\indent Based on the above, we can come up with a general formula for $\delta\epsilon_{\text{cl}}$:\,\footnote{The author wishes to thank an anonymous referee for a crucial observation regarding the general form of $\delta\epsilon_{\text{cl}}$.}
\small\begin{IEEEeqnarray}{l}
\delta\epsilon_{\text{cl}} = \frac{1}{\pi} \cdot \sum_{n = 1}^{\infty} \Bigg[\mathcal{A}_{n0}\left(p\right) \, \mathcal{J}^{2n - 2} + \mathcal{A}_{n1}\left(p\right) \, \mathcal{J}^{2n - 3} + \mathcal{A}_{n2}\left(p\right) \, \mathcal{J}^{2n - 4} + \ldots + \mathcal{A}_{n(2n - 2)}\left(p\right)\Bigg]\, e^{-n\mathcal{L}}, \qquad \label{GiantMagnon5}
\end{IEEEeqnarray}\normalsize
where the coefficients of all the negative powers of $\mathcal{J}$ vanish (e.g.\ $\mathcal{A}_{11} = \mathcal{A}_{12} = \ldots = 0$). In \cite{FloratosLinardopoulos14}, the coefficients $\mathcal{A}_{n0}$, $\mathcal{A}_{n1}$, $\mathcal{A}_{n2}$, have been called leading (L), next-to-leading/subleading (NL) and next-to-next-to-leading/next-to-subleading (NNL) respectively. The leading coefficients $\mathcal{A}_{10}$--$\mathcal{A}_{60}$ were determined by Klose and McLoughlin in 2008 \cite{KloseMcLoughlin08}:
\begin{IEEEeqnarray}{ll}
\delta\epsilon_{\text{cl}}\Big|_{\text{L}} = - \frac{4}{\pi} \, \sin^3\frac{p}{2} \, e^{-\mathcal{L}}\bigg[1 &+ 2\mathcal{L}^2\cos^2\frac{p}{2}\,e^{-\mathcal{L}} + 8\mathcal{L}^4\,\cos^4\frac{p}{2}\,e^{-2\mathcal{L}} + \frac{128}{3}\mathcal{L}^6\,\cos^6\frac{p}{2}\,e^{-3\mathcal{L}} + \nonumber \\[6pt]
& + \frac{800}{3}\mathcal{L}^8\,\cos^8\frac{p}{2}\,e^{-4\mathcal{L}} + \frac{9216}{5}\mathcal{L}^{10}\,\cos^{10}\frac{p}{2}\,e^{-5\mathcal{L}} + \ldots\bigg]. \label{GiantMagnon6}
\end{IEEEeqnarray}
In \cite{FloratosLinardopoulos14} all the classical coefficients $\mathcal{A}_{n0}$, $\mathcal{A}_{n1}$, $\mathcal{A}_{n2}$ were computed in closed forms. In the last section of this talk we are going to briefly review the method of \cite{FloratosLinardopoulos14} and present the results for the coefficients $\mathcal{A}_{n0}$, $\mathcal{A}_{n1}$, $\mathcal{A}_{n2}$.
\subsection[Closed-Form Expressions]{Closed-Form Expressions \label{Subsection:ClosedFormExpressions}}
\noindent The origins of the method that was introduced in \cite{FloratosLinardopoulos14} should be traced back to the 2010 paper of Georgiou and Savvidy \cite{GeorgiouSavvidy11}, who studied the dispersion relation of classical GKP strings that rotate rigidly inside AdS$_3$. Even though the exact classical expressions of all the conserved charges of the GKP strings are known in parametric form as functions of the string's angular velocity $\omega$, the corresponding anomalous dimensions have to be expressed solely in terms of the string's conserved charges, namely the angular momenta $J$ and $S$. Only in this way can they accommodate quantum corrections and be compared to the corresponding weak-coupling formulae, none of which has a parametric form in terms of $\omega$. The authors of \cite{GeorgiouSavvidy11} had the brilliant idea to investigate the inversion of the series that provided the conserved spin $S$ of the string, and then to use the "inverse spin function" that they had found in order to calculate various classical coefficients in the corresponding dispersion relation. \\[6pt]
\indent The Georgiou-Savvidy series inversion technique, was further refined in the paper \cite{FloratosGeorgiouLinardopoulos13}. By looking closely at the inversion algorithm and making it more systematic, the authors were able to spot certain regularities that could be described with a certain elementary function that is known as Lambert's W-function.\footnote{See appendix \ref{Appendix:LambertFunction} for the definition and some properties of Lambert's W-function.} This way it became possible to determine even more classical finite-size coefficients in the dispersion relation of the AdS$_3$ GKP string, but also to find the leading, subleading and next-to-subleading terms in the classical dispersion relation of the $\mathbb{R}\times\text{S}^2$ GKP string \eqref{GKP_String_II2}. \\[6pt]
\indent Based on what we have said about the connection of the GKP string in $\mathbb{R}\times\text{S}^2$ \eqref{GKP_String_II2} to the giant magnon, it's clear that a similar W-function description should be applicable to finite-size GMs as well. As we have already explained, the dispersion relation has to be expressed in terms of the various conserved charges of the system. Contrary to GKP strings however that have just two conserved charges (their energy $E$ and spin $S$ or $J$) depending on only one parameter (their angular velocity $\omega$), GMs have an additional conserved charge (their momentum/angular extent $\Delta\phi = p$) and an additional parameter, namely their linear velocity $v$. We are therefore led to a $3\times3$ system of equations that has the following general parametric form:
\begin{IEEEeqnarray}{c}
\mathcal{E} = d\left(a,x\right) \ln x + h\left(a,x\right) \label{GM_Charges1} \\[6pt]
\mathcal{J} = c\left(a,x\right) \ln x + b\left(a,x\right) \label{GM_Charges2} \\[6pt]
p = f\left(a,x\right) \ln x + g\left(a,x\right) \label{GM_Charges3}
\end{IEEEeqnarray}
and represents a much more challenging technical problem than the $2\times2$ system that we obtain in the case of GKP strings. In \eqref{GM_Charges1}--\eqref{GM_Charges3} we've defined $v \equiv \cos a$, $x = x\left(\omega,v\right)$, while $d, h, c, b, f, g$ are known power series of the variables $x$ and $a$. We've also defined $\mathcal{E} \equiv \pi E / \sqrt{\lambda}$ and $\mathcal{J} \equiv \pi J / \sqrt{\lambda}$. \\[6pt]
\indent Let us briefly sketch how the solution of the system \eqref{GM_Charges1}--\eqref{GM_Charges3} proceeds. More technical details can be found in the paper \cite{FloratosLinardopoulos14}. We first eliminate the logarithm from the last two equations, \eqref{GM_Charges2}--\eqref{GM_Charges3}. This leads to an equation $p = p\left(\mathcal{J}, a,x\right)$, where the momentum $p$ is a function of the conserved spin $\mathcal{J}$ and the variables $a$ and $x$. $p\left(\mathcal{J}, a,x\right)$ can be expanded in a double series in $a$ and $x$, which can subsequently be inverted for $a$, leading to an expression for $a = a\left(x, p, \mathcal{J}\right)$. If we plug $a\left(x, p, \mathcal{J}\right)$ back into the first two equations \eqref{GM_Charges1}--\eqref{GM_Charges2}, we obtain the following $2\times2$ system:
\begin{IEEEeqnarray}{c}
\mathcal{E} = d\left(x,p,\mathcal{J}\right) \ln x + h\left(x,p,\mathcal{J}\right) \\[6pt]
\mathcal{J} = c\left(x,p,\mathcal{J}\right) \ln x + b\left(x,p,\mathcal{J}\right),
\end{IEEEeqnarray}
which we may solve for $\mathcal{E} = \mathcal{E}\left(p,\mathcal{J}\right)$ along the lines of \cite{FloratosGeorgiouLinardopoulos13}. The result is: \\
\begin{IEEEeqnarray}{ll}
\mathcal{E} - \mathcal{J}\Big|_{\text{classical}} = \sin\frac{p}{2} &+ \frac{1}{4\mathcal{J}^2}\tan^2\frac{p}{2}\sin^3\frac{p}{2}\left[W + \frac{W^2}{2}\right] - \frac{1}{16\mathcal{J}^3}\tan^4\frac{p}{2}\sin^2\frac{p}{2}\bigg[\left(3\cos p + 2\right)W^2 + \nonumber \\[6pt]
& + \frac{1}{6}\left(5\cos p + 11\right)W^3\bigg] - \frac{1}{512\mathcal{J}^4}\tan^6\frac{p}{2}\sin\frac{p}{2}\Bigg\{\left(7\cos p - 3\right)^2\frac{W^2}{1 + W} - \nonumber \\[6pt]
& - \frac{1}{2}\left(25\cos2p -188\cos p -13\right)W^2 - \frac{1}{2}\left(47\cos2p + 196\cos p - 19\right)W^3 - \nonumber \\[6pt]
& - \frac{1}{3}\left(13\cos2p + 90\cos p + 137\right)W^4\Bigg\} + \ldots, \qquad \mathcal{J},\lambda \rightarrow \infty, \label{GiantMagnon7}
\end{IEEEeqnarray} \\[6pt]
where the argument of the W-function is $W_0\left(\pm 16 \mathcal{J}^2 \cot^2\left(p/2\right) e^{- 2\mathcal{J}\csc p/2 - 2}\right)$ in the $W_0$ branch (see appendix \ref{Appendix:LambertFunction}) and the sign $\pm$ refers to the elementary ($-$) and the doubled ($+$) region of GMs. \\[6pt]
\indent We may use the expansion of Lambert's W-function around the point $x = 0$ (given in equation \eqref{LambertSeries0} of appendix \ref{Appendix:LambertFunction}) to expand \eqref{GiantMagnon7} in a Taylor series around $\mathcal{J} \rightarrow \infty$. We will find that the second term in \eqref{GiantMagnon7} gives all the leading coefficients $\mathcal{A}_{n0}$ of \eqref{GiantMagnon5}, the third term gives all the next-to-leading coefficients $\mathcal{A}_{n1}$, while the fourth term in \eqref{GiantMagnon7} contains all the NNL coefficients $\mathcal{A}_{n2}$: \\
\begin{IEEEeqnarray}{l}
\text{leading: } \sum_{n = 1}^{\infty} \mathcal{A}_{n0}\left(p\right) \, \mathcal{J}^{2n - 2} \, e^{-n\mathcal{L}} = \frac{1}{4\mathcal{J}^2}\tan^2\frac{p}{2}\sin^3\frac{p}{2}\left[W + \frac{W^2}{2}\right] \label{GM_Leading}\\[6pt]
\text{subleading: } \sum_{n = 2}^{\infty} \mathcal{A}_{n1}\left(p\right) \, \mathcal{J}^{2n - 3} \, e^{-n\mathcal{L}} = - \frac{1}{16\mathcal{J}^3}\tan^4\frac{p}{2}\sin^2\frac{p}{2}\bigg[\left(3\cos p + 2\right)W^2 + \nonumber \\[6pt]
\hspace{10.2cm} + \frac{1}{6}\left(5\cos p + 11\right)W^3\bigg] \label{GM_Subleading} \\[6pt]
\text{next-to-subleading: } \sum_{n = 2}^{\infty} \mathcal{A}_{n2}\left(p\right) \, \mathcal{J}^{2n - 4} \, e^{-n\mathcal{L}} = - \frac{1}{512\mathcal{J}^4}\tan^6\frac{p}{2}\sin\frac{p}{2}\Bigg\{\left(7\cos p - 3\right)^2\frac{W^2}{1 + W} - \nonumber \\[6pt]
\hspace{3.5cm} - \frac{1}{2}\left(25\cos2p -188\cos p -13\right)W^2 - \frac{1}{2}\left(47\cos2p + 196\cos p - 19\right)W^3 - \nonumber \\[6pt]
\hspace{3.5cm} - \frac{1}{3}\left(13\cos2p + 90\cos p + 137\right)W^4\Bigg\}. \label{GM_Subsubleading}
\end{IEEEeqnarray} \\[6pt]
In \eqref{GM_Leading}--\eqref{GM_Subsubleading}, $\mathcal{A}_{10}$, $\mathcal{A}_{20}$, $\mathcal{A}_{21}$, $\mathcal{A}_{22}$ are the Arutyunov-Frolov-Zamaklar coefficients \eqref{GiantMagnon4}, while $\mathcal{A}_{10}$-$\mathcal{A}_{60}$ are the Klose-Mcloughlin coefficients \eqref{GiantMagnon6}. Comparing the coefficients $\mathcal{A}_{n0}$, $\mathcal{A}_{n1}$, $\mathcal{A}_{n2}$ that are found from the above formulas with those that have been computed with $\mathsf{Mathematica}$ in appendix B of reference \cite{FloratosLinardopoulos14}, we find that they completely agree.
\section[Discussion]{Discussion \label{Section:Discussion}}
\noindent Giant magnons are bosonic single-spin open strings that rotate in $\mathbb{R}\times\text{S}^2 \subset \text{AdS}_5\times\text{S}^5$. According to the AdS/CFT dictionary, giant magnons are the string theory duals of magnon excitations of $\mathcal{N} = 4$ SYM. Magnons appear in $\mathcal{N} = 4$ SYM when one considers the dilatation operator of the $\mathfrak{su}\left(2\right)$ sector of the theory, which is given by the Hamiltonian of the XXX$_{1/2}$ spin chain at one loop. Magnons and giant magnons are the elementary excitations of AdS/CFT, out of which all states in the theory are built. \\[6pt]
\indent The dispersion relation of magnons below the critical loop order is completely determined by the asymptotic Bethe ansatz (ABA). The ABA also fixes the dispersion relations of magnons and giant magnons at infinite size, i.e.\ when $J = \infty$. Below the critical loop order one has to calculate wrapping corrections on the weakly coupled side and classical and quantum finite-size corrections on the strongly coupled one. \\[6pt]
\indent In this talk we have presented a method to calculate classical finite-size corrections to the dispersion relation of giant magnons by using strings. Following \cite{FloratosLinardopoulos14}, we have inverted the expressions that give the conserved (linear and angular) momenta of GMs in terms of elliptic integrals. By plugging the resulting formulas into the expression of the conserved energy of GMs, we have obtained closed-form expressions for the leading \eqref{GM_Leading}, next-to-leading \eqref{GM_Subleading} and next-to-next-to-leading \eqref{GM_Subsubleading} series of finite-size corrections to the dispersion relation of giant magnons,
\begin{IEEEeqnarray}{c}
E - J = \epsilon_{\infty} + \sqrt{\lambda}\,\delta\epsilon_{\text{cl}} + \delta\epsilon_{1\text{-loop}} + \frac{1}{\sqrt{\lambda}}\, \delta\epsilon_{2\text{-loop}} + \ldots, \qquad J, \ \lambda \rightarrow \infty \nonumber \\
\delta\epsilon_{\text{cl}} = \frac{1}{\pi} \cdot \sum_{n = 1}^{\infty} \sum_{m = 0}^{2n - 2} \mathcal{A}_{nm}\left(p\right)\mathcal{J}^{2n - m - 2} e^{-2n\left(\mathcal{J}\csc\frac{p}{2} + 1\right)}, \quad \mathcal{J} \equiv \frac{\pi J}{\sqrt{\lambda}}, \nonumber
\end{IEEEeqnarray}
i.e.\ all the coefficients $\mathcal{A}_{n0}$, $\mathcal{A}_{n1}$, $\mathcal{A}_{n2}$. We may infer that all the higher order terms of the classical finite-size corrections $\delta\epsilon_{\text{cl}}$ will be given by some similar expression with Lambert W-functions. \\[6pt]
\indent We end this discussion with some thoughts on possible future projects. First it would be interesting to try to probe NkLO terms in the classical expansion $\delta\epsilon_{\text{cl}}$ by means of an algorithm, some iterative procedure or even a $\mathsf{Mathematica}$ program. This could pave the way for a better description (perhaps with a closed analytic formula) of the classical GM spectrum at finite size. With the new analytic tool that we've presented here, we could also envisage revisiting some more complicated spectral problems for the GM, such as the computation of quantum corrections $\delta\epsilon_{n\text{-loop}}\,$, or wrapping corrections at weak coupling. \\[6pt]
\indent Another very appealing prospect would be to try to make contact with other spectral methods that account for wrapping effects (e.g.\ L\"{u}scher corrections, TBA/Y-system/QSC). Perhaps a more powerful spectral technique could result from their combination with the method that is presented here. Many other generalizations of our work can be thought of. For example, dispersion relations in ABJM theory,\footnote{For an application of the W-function to the dispersion relation of strings that rotate inside AdS$_4\times\mathbb{CP}^3$, see \cite{DimovMladenovRashkov14}.} AdS spacetime, deformed backgrounds, spiky strings, M2-branes,\footnote{Along the lines of \cite{AxenidesFloratosLinardopoulos13a} for example.} the computation of correlation functions, etc.\ could all afford a W-function parametrization.
\acknowledgments
\noindent The author is grateful to the organizers of the 2014 Corfu Summer Institute and especially to professor George Zoupanos for the invitation to participate and present a talk in the very exciting Workshop on Quantum Fields and Strings. Minos Axenides is essentially the person who made all the financial arrangements that allowed the author to travel to Corfu. The author is especially grateful to him and also to professor Emmanuel Floratos for their advices and assistance. \\[6pt]
\indent The author would also like to thank professors Ioannis Bakas, Joseph Minahan, Konstantinos Sfetsos and Nikolaos Tetradis for illuminating discussions about the work herein presented. The author is indebted to Minos Axenides, George Georgiou and Stam Nicolis for their help with the manuscript. Last but not least the author wishes to thank all the participants of the Workshop with whom he shared many interesting discussions. \\[6pt]
\indent All the plots of this paper have been drawn with $\mathsf{Mathematica}$. The research of G.L. at N.C.S.R. "Demokritos" is supported by the General Secretariat for Research and Technology of Greece and from the European Regional Development Fund MIS-448332-ORASY (NSRF 2007--13 ACTION, KRIPIS).
\begin{figure}
\begin{center}
\includegraphics[scale=0.4]{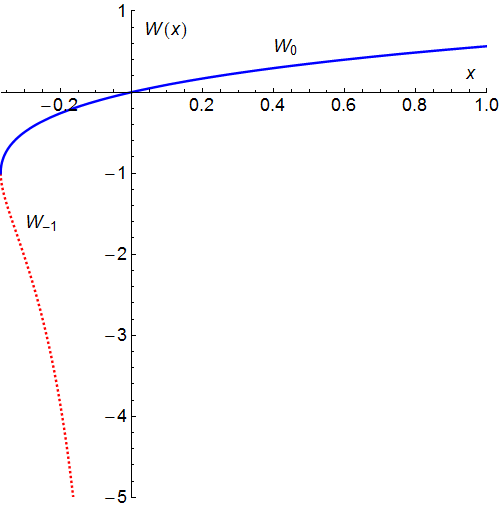}
\caption{The two real branches of Lambert's W-function.} \label{Graph:LambertFunction}
\end{center}
\end{figure}
\appendix
\section[Lambert's W-Function]{Lambert's W-Function \label{Appendix:LambertFunction}}
\noindent Lambert's W-function is defined by the following implicit formula:
\begin{IEEEeqnarray}{c}
W\left(z\right)\,e^{W\left(z\right)} = z \Leftrightarrow W\left(z\,e^z\right) = z. \label{LambertDefinition}
\end{IEEEeqnarray}
The W-function has two real branches, $W_0\left(x\right)$ for $x \in \left[-e^{-1},\infty\right)$ and $W_{-1}\left(x\right)$ for $x \in \left[-e^{-1},0\right]$, that have been drawn in figure \ref{Graph:LambertFunction}. The branch point is $\left(-e^{-1},-1\right)$. The Taylor series around $x = 0$, in the $W_0$ branch is \cite{CorlessGonnetHareJeffreyKnuth96}: \\
\begin{IEEEeqnarray}{l}
W_0\left(x\right) = \sum_{n = 0}^\infty \left(-1\right)^n\frac{\left(n+1\right)^n}{\left(n+1\right)!}\cdot x^{n+1} = \sum_{n = 1}^\infty \left(-1\right)^{n + 1} \frac{n^{n - 1}}{n!}\cdot x^n\,, \qquad \left|x\right| \leq e^{-1}. \label{LambertSeries0}
\end{IEEEeqnarray} \\[6pt]
The W-function also provides the limiting value of the tetration $x^{x^{x^{\ldots}}}$: \\
\begin{IEEEeqnarray}{c}
x^{x^{x^{\ldots}}} = \tensor[^\infty]{\left(x^{z}\right)}{} = \frac{W\left(-\ln x\right)}{-\ln x}.
\end{IEEEeqnarray}
\bibliographystyle{JHEP}
\bibliography{Bibliography}
\end{document}